\documentclass[aps,prl,epsf,twocolumn,showpacs,floatfix,preprintnumbers,amsmath,amssymb,superscriptaddress]{revtex4}
\usepackage{graphicx}

\input{epsf}
\epsfclipon

\begin{document} 

\title{Minimal model for active nematics: 
quasi-long-range order and giant fluctuations}

\author{Hugues Chat\'e}
\affiliation{CEA -- Service de Physique de l'Etat Condens\'e, Centre d'Etudes de Saclay, 91191 Gif-sur-Yvette, France}

\author{Francesco Ginelli}
\affiliation{CEA -- Service de Physique de l'Etat Condens\'e, Centre d'Etudes de Saclay, 91191 Gif-sur-Yvette, France}

\author{Ra\'ul Montagne}
\affiliation{Departamento de F\'{\i}sica, Universidade Federal de Pernambuco,
Cidade Universit\'aria, 50670-901, Recife-PE, Brazil}

\begin{abstract}
We propose a minimal microscopic model for active nematic particles 
similar in spirit to the Vicsek model for self-propelled polar particles.
In two dimensions, we show that this model
exhibits a Kosterlitz-Thouless-like
transition to quasi-long-range orientational order and that 
in this non-equilibrium context, the ordered phase is characterized by giant
density fluctuations, in agreement with the predictions of 
Ramaswamy {\it et al.} [Europhys. Lett. {\bf 62}, 196 (2003)].
\end{abstract}

\maketitle

%% say somewhere that we do not take hydrodynamic interactions into account
%% disclination unbinding
%% phase separation and number fluctuations

Over the last decade or so, physicists have been looking for common, 
possibly universal, features of the collective motion of animals, bacteria,
cells, molecular motors, as well as driven granular objects \cite{REVIEW,REVIEW-IGOR}. 
Among the emergent properties of these groups of 
``active'' or self-propelled particles (SPP),
distinctively out-of-equilibrium features have been found, such as the 
existence of long-range orientational order in two-dimensional 
``ferromagnetic'' flocks of polar SPP \cite{VICSEK,TT}. 

Another set of striking intrinsically non-equilibrium properties
have recently been predicted by Ramaswamy and co-workers \cite{SR,RST,REVIEW}.
They considered in particular the case of apolar but
oriented SPP and argued that such ``active nematics'' 
should differ dramatically from the 
usual (equilibrium) case \cite{CRYSTALS}. 
In particular, their approach, based on the analysis
of hydrodynamic equations derived from symmetry arguments,
predicts that giant density fluctuations arise in the ordered phase of 
such media.
In \cite{REVIEW}, it is also hinted at the possibility of true long-range order
and of a different isotropic-nematic transition out-of-equilibrium, but
no definitive statement is offered. Resolving these issues is nevertheless
crucial, especially in view of the predicted giant density fluctuations,
and all the more so since, in polar SPP, the transition to true 
long-range order was shown to be discontinuous \cite{BOID-PRL}.

In spite of the above-mentioned current surge of activity in non-equilibrium
systems, the
giant density fluctuations predicted by Ramaswamy and co-workers
have not been observed so far, and the nature of the nematically-ordered
phase and of the transition leading to it have not been elucidated.
Experimentally, relevant systems such as
colonies of elongated cells \cite{GRULER} and ensembles of 
rod-like objects driven by vibration \cite{NEICU,GALANIS,RAMA-EXP}
have been studied, but with other issues at stake. 
On the theoretical side, no microscopic model has been proposed \cite{NOTE}.
In this Letter, we fill this gap, confirm for the first time 
the predictions of Ramaswamy {\it et al.}, and investigate the 
nature of the isotropic-nematic transition in driven systems.
We introduce a minimal model for active nematics, and 
show numerically that its isotropic-nematic transition in two
space dimensions does not differ significantly from the equilibrium case:
only quasi-long-range (QLRO) order is attained, with scaling laws 
compatible with those of the Kosterlitz-Thouless (KT) transition.
Nevertheless, giant density fluctuations are clearly observed in the 
ordered phase: the standard deviation $\Delta n$ of $n$, the average number 
of particles in a given sub-system, is proportional to $n$ and not to
$\sqrt{n}$ as expected in equilibrium.

Our model is similar in spirit to the Vicsek model for polar SPP
\cite{VICSEK}. In a typical driven-overdamped dynamics,
identical pointwise particles
move synchronously at discrete timesteps $\Delta t$
by a fixed distance $v_0 \Delta t$.
In two space dimensions and for uniaxial nematics
---the case to which we restrict ourselves in the following---,
each particle $j$ is endowed with an orientation $\theta_j$
and moves along $\theta_j$ or $\theta_j+\pi$ with
equal probabilities. At every timestep,
$\theta_j^{t+1}$ is given by $\Theta({\bf Q}_j^t)$, the direction of the 
first eigenvector of the {\it local} tensorial traceless order parameter 
\begin{equation}
{\bf Q}_j = \left(\begin{array}{cc}
\langle\cos^2\theta_k\rangle-\frac{1}{2} & \langle\cos\theta_k\sin\theta_k\rangle\\
-\langle\cos\theta_k\sin\theta_k\rangle & \langle\sin^2\theta_k\rangle-\frac{1}{2} \end{array} \right)
\label{eq:Q}
\end{equation}
where the average is taken over all particles $k$ 
within the interaction range $r_{\rm 0}=1>v_0 \Delta t$, 
including particle $j$ (in this paper, we use $v_0 \Delta t=0.3$).
As for the Vicsek model, disorder arises from the addition of
a random angle to this newly calculated orientation, and we have finally:
\begin{equation}
\theta_j^{t+\Delta t} = \Theta({\bf Q}_j^t) + \sigma \, \xi_j^t  \;,
\label{eq:Dynamics}
\end{equation}
where $\xi_j^t$ is a delta-correlated white noise 
($\xi\in [-\frac{\pi}{2},\frac{\pi}{2}]$).
The interaction introduces a tendency to align (nematically)
with neighboring particles, so that two simple limits arise: 
Complete orientational order settles in the absence of noise, whereas
particles perform random walks for maximal noise ($\sigma=1$).
We first characterize the transition that necessarily 
lies in between these two regimes. To this aim, we
calculate the total order parameter ${\bf Q}(N,L)$
measured for $N$ particles in a square domain of linear size $L$ with 
periodic boundary conditions. We use in particular the scalar order parameter
\begin{equation}
S=2\sqrt{
(\langle\cos^2\theta\rangle-\frac{1}{2})^2+
\langle\cos\theta\sin\theta\rangle^2} \;.
\end{equation}
which is equal to twice the (positive) eigenvalue of ${\bf Q}$, 
so that $S=1$ for perfect orientational order, 
and $S=0$ for complete disorder. 
Starting from random positions and orientations, 
$S$ typically grows in time and eventually reaches a statistically 
stationary state characterized by a well-defined distribution function of mean
$\langle S \rangle$.

\begin{figure}
\centerline{
\epsfxsize=8.6cm
\epsffile{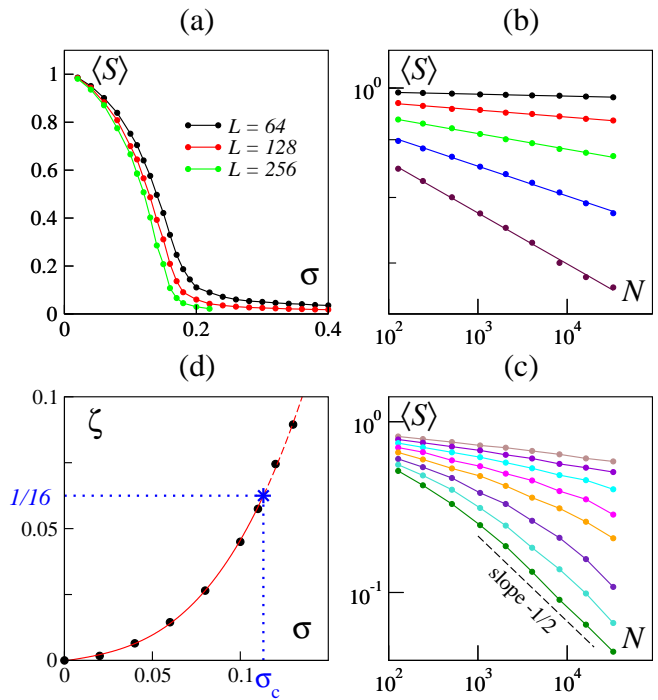}
}
\caption{(Color online) 
Transition to nematic order at density $\rho=N/L^2=\frac{1}{2}$.
(a): (time-averaged) scalar order parameter $\langle S\rangle$ 
vs noise strength $\sigma$ for various sizes.
(b): $\langle S\rangle$ vs $N$ at $\sigma=0.02$, 0.04, 0.06, 0.08, 0.1 
from top to bottom; the lines are fitted power-laws. Each point represents an
average over $10^6-10^7$ timesteps after transients.
(c): same as (b) but for larger noise values: $\sigma=0.11$, 0.12, 0.13,
0.14, 0.15, 0.16, 0.17, 0.18. For large $\sigma$ values these curves crossover 
to $1/\sqrt{N}$ decay.
(d): exponents $\zeta$ extracted from the power-laws shown in (b) and (c)
vs $\sigma$. The line linking the symbols is a (cubic) fit. 
For the last 2 points, the exponent is only effective, as 
it was estimated from small $N$ values, before signs of 
crossover appear. The dotted lines indicate the threshold 
$\sigma_{\rm c}=0.113(5)$ estimated from the condition 
$\zeta(\sigma_{\rm c})=\frac{1}{16}$ (see text).
}
\label{fig1}
\end{figure}

Varying the noise intensity $\sigma$, we observe
a continuous change of $\langle S \rangle$.
Increasing system size at fixed density $\rho=N/L^2$, 
the curves $S(N)$ vs $\sigma$  reveal sharper transitions for larger sizes,
but they do {\it not} 
cross each other (Fig.~\ref{fig1}a). Figure~\ref{fig1}b shows that
$ \langle S(N)\rangle \sim N^{-\zeta(\sigma)}$.
Increasing the noise 
strength from $\sigma=0$ towards the transition zone,
$\zeta(\sigma)$ increases from zero to take rather small values. 
Sufficiently deep in the transition zone, the effective exponent 
$\bar{\zeta}(\sigma, N)=-\frac{{\rm d}\ln S}{{\rm d}\ln N}$
can be observed to cross over, as $N \to \infty$,
from these small values towards 
$\frac{1}{2}$ (Fig.~\ref{fig1}c), the value observed 
at larger $\sigma$ and characteristic of a completely disordered phase.
All these observations are in qualitative agreement with an equilibrium
KT transition \cite{KT}, 
and signal that only QLRO order is present in the 
ordered phase. At the quantitative level, the location of the KT 
transition point in an equilibrium system 
is characterized by $\zeta=\frac{1}{16}$. 
Our data is consistent with this: for $\rho=\frac{1}{2}$,
we find $\zeta\simeq\frac{1}{16}$ at
$\sigma_{\rm c}\simeq 0.113(5)$, and for larger noise values 
$\bar{\zeta}(\sigma, N)$ show signs
of crossing over to $\frac{1}{2}$ (i.e. one is in the disordered phase).

\begin{figure}
\centerline{
\epsfxsize=8.6cm
\epsffile{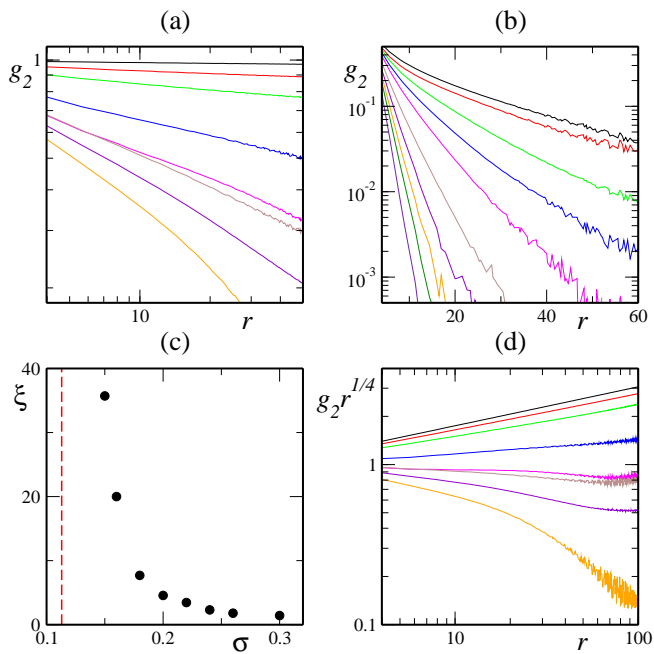}
}
\caption{(Color online) 
Orientational correlation function $g_2(r)$ as calculated from the 
orientation field coarse-grained over boxes of linear size 4
($\rho=\frac{1}{2}$, time-average over $10^6$-$10^7$ timesteps after transients 
for each run, $L=256$ in (a,d) and $L=128$ in (b)). 
(a) in log-log scales at small noise values $\sigma=0.02$, 0.04, 0.06, 
0.08, 0.10, 0.11, 0.12, and 0.13 from top to bottom.
(b) in lin-log scales for $\sigma=0.14$ to 0.3 from top to bottom.
(c) variation of the 
correlation length $\xi$ extracted from the exponential tails in (b)
as the transition point is approached (dashed line at the estimated threshold
$\sigma_{\rm c}=0.113$).
(d) same as in (a), data multiplied by the expected exponent at threshold 
$\eta_{\rm c}=\frac{1}{4}$:
around the estimated threshold value $\sigma_{\rm c}=0.113(5)$, 
the curves are flat and straight. 
}
\label{fig2}
\end{figure}

These observations are strengthened by the study of 
the orientational spatial correlation functions 
$g_{2m}(r) = \langle\cos[2m(\theta(0)-\theta(r))]\rangle$ 
(here $m$ is integer and averages are taken both in space and time). 
They decay algebraically at low noise values
($g_{2m}(r) \sim r^{-m^2\eta}$ with $\eta$ also increasing with $\sigma$)
and exponentially in the disordered phase (with a diverging correlation 
length $\xi$ as $\sigma$ approaches the critical point) (Fig.~\ref{fig2}a,b). 
At a quantitative level, one expects that  in equilibrium 
$\eta(\sigma)=4\zeta(\sigma)$ in the ordered phase \cite{KT}. 
This is roughly borne out of our data,
even though good estimates of the correlation functions are difficult to 
obtain close to the transition. 
In the disordered phase, in particular, the expected divergence of $\xi$
is observed (Fig.~\ref{fig2}c) but not in its expected functional dependence
as $\xi$ can only be safely estimated rather far away
from threshold. Despite these difficulties, we can
check that the critical noise level determined above is consistent with our
correlation function data (Fig.~\ref{fig2}d).

The above results are characteristic of QLRO order and of a 
KT-like phase transition. Further preliminary results \cite{TBP} 
show that the disclination unbinding mechanism
characteristic of the KT transition \cite{KT2} 
is also at work in our non-equilibrium
context. All this indicates the proper but costly methodology 
to locate the critical noise level $\sigma_{\rm c}$. We have not, so far,
used this protocol extensively, but ongoing simulations indicate that the
critical line may scale like $\sigma_{\rm c} \sim \rho^\frac{3}{4}$ and extend 
to arbitrarily small densities and noise levels.

\begin{figure}
\centerline{
\epsfxsize=8cm
\epsffile{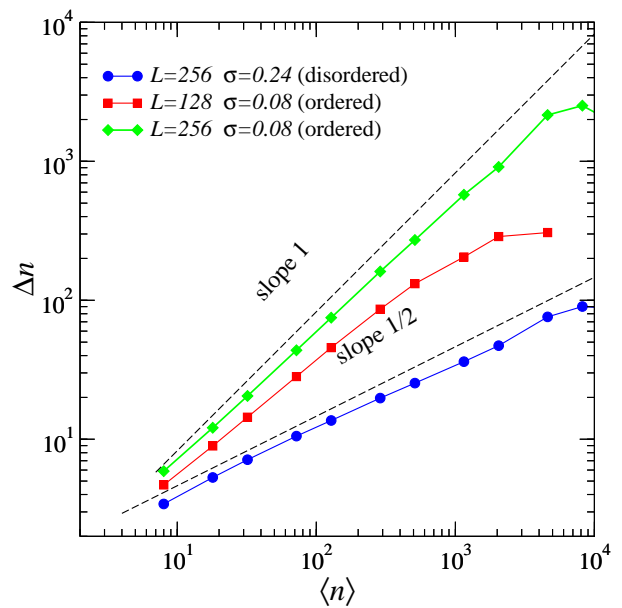}
}
\caption{(Color online)
Root mean square density fluctuations $\Delta n$ 
in square boxes containing $\langle n\rangle$ particles on average 
(linear size $\ell=\sqrt{n/\rho}$, $\rho=\frac{1}{2}$, 
time-average over $10^6$-$10^7$ timesteps after transients). 
The saturation at large $\langle n\rangle$ occurs when $\ell\simeq L$.
Top two curves: giant fluctuations in the ordered phase at 
2 different system sizes.
Bottom curve: normal fluctuations in the disordered phase.}
\label{fig3}
\end{figure}

\begin{figure}
\centerline{
\epsfxsize=8.6cm
\epsffile{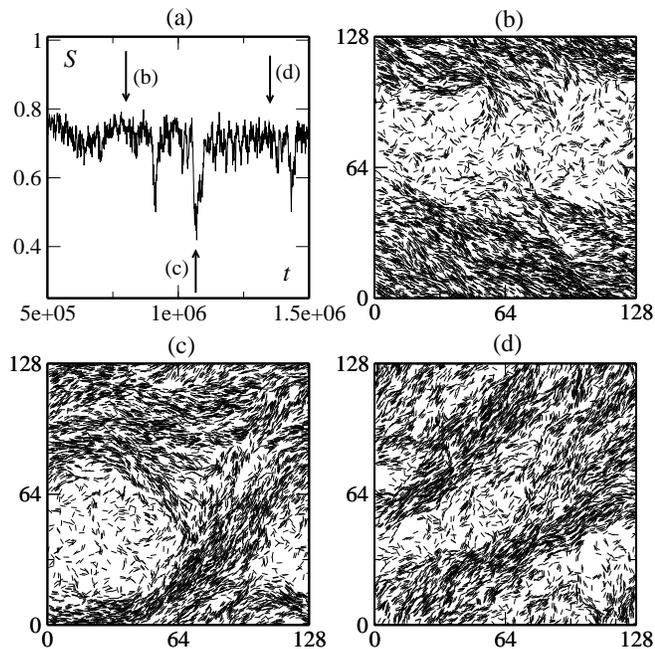}
}
\caption{(Color online) Typical time series of $S$ in the ordered phase.
($\rho=\frac{1}{2}$, $L=128$, $\sigma=0.1$). 
Note the large excursions to small $S$ values which make time-averaging 
difficult. 
(b-d) snapshots taken during the run shown in (a) at the times indicated there.
For each particle, a small segment (of arbitrary length) centered 
on its position and aligned on its orientation is drawn.
(b) and (d) are typical ordered states, while (c) represents 
the more disordered episodes
when the macroscopic structure changes orientation.
}
\label{fig4}
\end{figure}

In spite of the equilibrium-like properties of the transition to QLRO,
the non-equilibrium character of the problem is manifests itself in strong
density fluctuations, as predicted by Ramaswamy {\it et al.} \cite{RST}:
we measured, in the ordered phase, the density fluctuations in square
boxes of linear size $\ell$ embedded in a square domain of linear size $L$. 
These boxes contain, on average, 
$\langle n \rangle=\rho \ell^2$ particles.
As long as $\ell < L$, $\Delta n$,
the rms of the fluctuations of $n$, scales 
linearly with $\langle n \rangle$ (and {\it not} $\sqrt{n}$), 
in agreement with \cite{RST} (Fig.~\ref{fig3}).
These giant number fluctuations are the statistical consequence of
the complex, coupled, spatiotemporal dynamics of density and orientation
in the system. After transients,
low- and high-density regions emerge with the highly populated domains
taking the form of bands
inside which nematic order is strong (Fig.~\ref{fig4}). These bands
evolve (move, split, merge, dissolve and form again along a new direction)
over very long timescales. Typically,
however, a single band is present at 
any given time, independently of the system size, and its 
characteristic evolution time grows with system size.
Thus, contrary to what has been observed in the 
polar SPP case \cite{BOID-PRL}, 
these structures have no well-defined length- or time-scales.
Ongoing work aims at quantifying these statements \cite{TBP}.
We now comment on our results. In our minimal model, the density and 
orientation fields are coupled intimately. Whether
the mechanism put forward in \cite{RST} ---namely that the current in 
the conservation equation for the concentration has a contribution
proportional to $(\partial_y\theta,\partial_x\theta)$--- is actually present
here remains to be seen explicitly, but we believe that this is the
case because the general symmetry arguments invoked there
must apply. As a matter of fact, in our model, intrinsically non-equilibrium
features are still observed when the displacement of the particles 
is not made along their orientation 
---the ``natural'' case if one has in mind particles with
an elongated physical shape---, but,
for instance, {\it perpendicularly} to their axis (not shown).
On the other hand, displacing them randomly
along one of the four directions defined by adding 
multiples of $\frac{\pi}{2}$ to their current angle
yields normal fluctuations ($\Delta n \sim \sqrt{n}$) and 
no segregation (not shown).
This is not surprising since then particles effectively perform
random walks on scales larger than the elementary displacement. 
This case is thus equivalent to 
strictly decoupling density fluctuations 
from the orientation field
by letting particles be non-interacting random walkers.
Apart from such ``equilibrium'' cases, it seems that any coupling
generically triggers giant fluctuations, so that they appear as a robust 
feature, as implicitly implied by the general arguments developed 
in \cite{RST}.
Whereas giant number fluctuations are easily observed, 
we have not been able to measure the 
slow decay of tagged-particle velocity autocorrelations also predicted in 
\cite{RST}. We believe this is because the only significant motion in our model,
apart from the microscopic random displacements, is due to the very
slow dynamics of the high-density bands (Fig.~\ref{fig4}). 
Thus, we expect such an effect
to be only observable on timescales so large that they are not easily accessible.

The giant number fluctuations taken as the signature of the 
``non-equilibriumness'' of the system are tantamount to the formation
of the high-density ordered band described above (Fig.~\ref{fig4}).
Whether this is ``true'' macroscopic phase separation is thus a key question,
which we just learnt to be the subject of \cite{RAMA-PREPRINT}. 
Here we have shown that despite this spectacular phenomenon, 
the phase transition is similar, as far as we can tell numerically, to
the equilibrium one, with an ordered phase characterized 
by QLRO only. This has to be paralleled to the case of polar SPP,
where true long-range order is ascertained \cite{TT}, and the transition
is discontinuous (i.e. first-order like) \cite{BOID-PRL}. 
At equilibrium, the Mermin-Wagner theorem states that true LRO cannot arise 
from the spontaneous breaking of a continuous symmetry in two dimensions
\cite{MW}.
Out of equilibrium, this constraint disappears \cite{TT,REVIEW}, and it 
is quite interesting to notice that polar and nematic SPP behave differently.
Note also that
for polar particles the ordered phase is {\it not} density-homogeneous, 
but typically consists of well-defined solitary bands with high density and 
strong order moving in a low-density disordered background \cite{BOID-PRL}.
These bands, which can appear in arbitrary numbers depending on the 
system geometry, are very different from the single, fluctuating,
splitting and merging object described here in Fig.~\ref{fig4}. 
Thus, combining the present conclusions and the results obtained recently on 
polar SPP, we see clearly how the non-equilibrium nature of these nonlinear
driven systems can emerge differently at the collective level. \\
To summarize, we have introduced a minimal model for active nematic particles
in which the density and orientation fields are coupled in a natural way.
This non-equilibrium model exhibits a KT-like phase transition 
to a nematically-ordered phase in which giant density fluctuations arise.
This constitutes a first non-trivial confirmation of the intrinsically 
non-equilibrium properties predicted in \cite{RST} and calls for further 
experimental studies, either with assemblies of granular elongated
particles or, better, in biological systems where such effects could play
an important role. In view of the notorious difficulties encountered to
decide about similar issues in the case of polar particles, 
our results will need to be confirmed by a proper renormalisation group
analysis.
On the modeling side, finally, our approach can easily be extended
to other space dimensions and/or to more complex types of interactions.
Future work will explore these issues, in particular three-dimensional
systems and the nature of tetratic order out of equilibrium.

We thank Sriram Ramaswamy for fruitful exchanges and the communication of
his recent results \cite{RAMA-PREPRINT}.


\begin{thebibliography}{99}
\bibitem{NOTE} After sending a draft of this paper to S. Ramaswamy,
we learnt of his ongoing work on the connection between phase separation
and giant fluctuations in active nematics \protect\cite{RAMA-PREPRINT}. 
To obtain these, he and S. Mishra
use an ad-hoc model in which lattice gas particles are influenced by an
independent equilibrium nematic system via a mechanism directly inspired 
by the analysis of \protect\cite{RST}.
\end{thebibliography}
\end{document}